\documentclass[useAMS,usegraphicx,usenatbib]{mn2e_arxiv}
\usepackage{amsmath}
\usepackage{fmtcount}
\usepackage[caption=false]{subfig}

\title[Globular cluster pulsar population]
{Constraining the luminosity function parameters and population size of radio
pulsars in globular clusters}
\author[Chennamangalam, Lorimer, Mandel and Bagchi]
{Jayanth Chennamangalam$^1$\thanks{E-Mail: jchennam@mix.wvu.edu},
D. R. Lorimer$^{1,2}$, Ilya Mandel$^{3}$
\newauthor
and Manjari Bagchi$^1$\\
$^1$ Department of Physics, West Virginia University, PO Box 6315, Morgantown, WV 26506, USA\\
$^2$ NRAO, Green Bank Observatory, PO Box 2, Green Bank, WV 24944, USA\\
$^3$ School of Physics and Astronomy, University of Birmingham, Edgbaston, Birmingham B15 2TT, UK}

\begin{document}

\date{}

\maketitle

\begin{abstract}
Studies of the Galactic population of radio pulsars have shown that their
luminosity distribution appears to be log-normal in form. We investigate some of the
consequences that occur when one applies this functional form to populations of
pulsars in globular clusters. We use Bayesian methods to explore constraints on
the mean and standard deviation of the luminosity function, as well as the total number of
pulsars, given an observed sample of pulsars down to some limiting
flux density, accounting for measurements of flux densities of
individual pulsars as well as diffuse emission from the direction of the cluster.
We apply our analysis to Terzan~5, 47~Tucanae and M~28, and demonstrate, under
reasonable assumptions, that the number of potentially observable pulsars should
be within 95\% credible intervals of $147^{+112}_{-65}$, $83^{+54}_{-35}$
and $100^{+91}_{-52}$, respectively. Beaming considerations would increase the
true population size by approximately a factor of two. Using non-informative
priors, however, the constraints are not tight due to the paucity and
quality of flux density measurements. Future cluster pulsar
discoveries and improved flux density measurements would allow this method to
be used to more accurately constrain the luminosity function, and to compare the luminosity
function between different clusters.
\end{abstract}


\section{Introduction}
Globular clusters are spherical collections of stars located throughout the
haloes of galaxies. Once thought to be composed entirely of old metal-poor
population II stars, they are now also believed to form during interactions
or collisions of galaxies, therefore containing younger stars having higher
metallicities \citep[see][]{zep03}. The total masses of globular clusters range up
to the order of $10^6 M_{\sun}$ \citep[see][]{mey97}, and core stellar number
densities reach $10^6$ pc$^{-3}$. The high core densities lead to dynamical interactions
between stellar systems that are found less commonly in the Galactic plane.
For example, globular clusters favour the formation of low-mass X-ray binaries
(LMXBs) that are believed to be the progenitors of millisecond pulsars
\citep[MSPs;][]{alp82,arc09}, and hence, the fraction of MSPs among
all pulsars in globular clusters is much larger than that in the Galactic field
($\sim$97\% versus $\sim$11\%). In addition, the binary MSPs in globular clusters tend to have
higher eccentricities compared to their field counterparts, due to
exchange or fly-by encounters. MSPs, due to their formation history, can
be considered long-lived tracers of LMXBs, and therefore, constraints on the
MSP content of globular clusters provide unique insights into binary evolution
and the integrated dynamical history of globular clusters, while determining
the radio luminosity function of these pulsars helps shed light on the radio
emission mechanism in action in these compact objects.

Pulsar searches of globular clusters have yielded impressive returns in recent
years \citep[see][]{cam05}, with currently 144 pulsars known in 28
clusters\footnote{See Paulo Freire's globular cluster pulsar catalogue at
http://www.naic.edu/$\sim$pfreire/GCpsr.html}. Of these, three clusters,
Terzan 5, 47 Tucanae and M 28 are known to harbour more than
10 pulsars each, the most populous being Terzan 5 with 34 (Ransom S. M.,
private communication). In this paper, we describe a Bayesian method that we
have developed to compute an estimate of the true number of pulsars in a given
cluster, given an observed population. There have been many attempts to constrain the
population size of pulsars in all globular clusters in the Galaxy
\citep*[see, for example,][]{kul90b,bag11}. This work is different in that it treats
clusters individually instead of dealing with the total population. Bayesian
approaches to constrain the pulsar population of individual globular clusters
have been used specifically for the case of young (non-recycled) pulsars by
\citet{boy11}. This work focuses on the entire radio pulsar content of the
cluster -- the majority of which is made up of old (recycled) pulsars -- and
additionally, attempts to constrain luminosity function parameters jointly with
population size.

\citet{fau06} have shown that the luminosity distribution of
non-recycled pulsars in the Galactic field appears to be
log-normal in form. More recently, \citet{bag11}
have verified that the observed luminosities of recycled pulsars in globular
clusters are consistent with this result. Assuming, therefore, that there is no significant
difference between the nature of Galactic and cluster populations, we
investigate some of the consequences that occur when one applies this functional
form to populations of pulsars in individual globular clusters.

For a log-normal (base-10) distribution of pulsar luminosities, the luminosity
function is given by
\begin{equation}
f(\log L) = \frac{1}{\sigma \sqrt{2\pi}} e^{-\frac{(\log L - \mu)^2}{2 \sigma^2}},
\end{equation}
where $L$ is the luminosity in mJy kpc$^2$, $\mu$
is the mean and $\sigma$ is the standard deviation of the distribution. We are
interested in the situation where we observe $n$ pulsars with luminosities
above some limiting luminosity $L_{\rm min}$. Given this sample of pulsars, we ask
what constraints we can place on their luminosity function parameters, in
addition to the potentially observable population size $N$ (that is, the population
of pulsars beaming towards the Earth). Another way of thinking
about this problem is that there is a family of luminosity function parameters and population sizes
that is consistent with an observation of $n$ pulsars above
the luminosity limit of the survey, and we are analyzing the posterior
probabilities of different members of this family given the observations.

This paper is organized as follows: In \S\ref{sec_bayesian}, we describe our
technique. In \S\ref{sec_apps}, we apply the technique
to observations of a few globular clusters to determine the constraints on the
luminosity function parameters and population size. Later, we refine our
results using \emph{a priori} information on the luminosity function parameters
to get a better estimate of the number of pulsars in those clusters. A summary
and our conclusions are presented in \S\ref{sec_conclusions}.

\section{Bayesian parameter estimation} \label{sec_bayesian}

Bayes' theorem \citep[see][]{wal03,gre05}, for
the purpose of parameter estimation, can be stated mathematically as
\begin{equation}
p(\boldsymbol{\theta} | D, M) = \frac{p(D | \boldsymbol{\theta}, M) p(\boldsymbol{\theta} | M)}{p(D | M)},
\end{equation}
where $\boldsymbol{\theta}$ is a set of parameters, $D$ is some data and $M$ is
a model describing the parameters. In this notation, $p(\boldsymbol{\theta} | D, M)$
represents the probability of obtaining a set of parameter values given the
data and the model, and is termed the \emph{joint posterior probability density}.
Similarly, $p(D | \boldsymbol{\theta}, M)$ is the probability of having obtained the
observed data, given the parameter values and the model, and is termed the
\emph{likelihood}, and $p(\boldsymbol{\theta} | M)$, the \emph{a priori}
probability dictated by the model, is termed the \emph{prior probability density}.
The denominator, $p(D | M)$ is called the \emph{evidence}, and is just a
normalizing factor that can be dropped since we are only interested in
relative probabilities, thereby giving
\begin{equation}
p(\boldsymbol{\theta} | D, M) \propto p(D | \boldsymbol{\theta}, M) p(\boldsymbol{\theta} | M).
\end{equation}
In this paper, we use Bayes' theorem to find the joint posterior probability
density functions of the model parameters $\mu$, $\sigma$ and $N$ given
some data. In our case, the data are the individual pulsar flux densities
that we call $\{S_i\}$, the observed number of pulsars, $n$, and the total diffuse
flux density of the cluster, $S_{\rm obs}$.

Luminosity is a property intrinsic to pulsars, while flux density is the
corresponding observable. The relationship between the two quantities is
given by
\begin{equation}
\label{eq_luminosity}
L = \frac{4 \pi r^2}{\delta} \sin^2 \left(\frac{\rho}{2}\right) \int_{\nu_1}^{\nu_2} S_{\rm mean}(\nu)~d\nu,
\end{equation}
where $r$ is the distance to the pulsar, $\delta$ is the pulse duty cycle,
$\rho$ is the radius of the pulsar emission cone, $S_{\rm mean}(\nu)$ is the
mean flux density of the pulsar as a function of observing frequency and
$\nu_1$ and $\nu_2$ are the bounds of the frequency range over which the
pulsar is observed \citep[see][]{lor05}. Due to the uncertainty associated with
the beam geometry, the values of $\delta$ and $\rho$ are not generally reliable
for luminosity calculations. Therefore, we use a simplified model of the luminosity,
the `pseudo-luminosity', that is defined as $L_\nu = S_\nu~r^2$ at a given frequency $\nu$
(the subscript $\nu$ on $L$ and $S$ will be dropped for the rest of the paper). As can be inferred
from Equation~(\ref{eq_luminosity}) and the aforementioned pseudo-luminosity equation,
the luminosity function is inevitably corrupted by uncertainties in distance.
To mitigate this, we decided to perform our analysis initially
in terms of the measured flux densities, and later, use a model of distance
uncertainty to convert our results to the luminosity domain.
We take the distance to all pulsars in a globular cluster to be the same. The
log-normal in luminosity can then alternatively be written in terms of the flux density. The
probability of detecting a pulsar with flux
density $S$ in the range $\log S$ to $\log S + d(\log S)$ is then given by a
log-normal in $S$ as
\begin{equation}
p(\log S)~d(\log S) = \frac{1}{\sigma_S \sqrt{2 \pi}}e^{-\frac{(\log S - \mu_S)^2}{2 \sigma_S^2}}~d(\log S),
\end{equation}
where $S$ is in mJy, and $\mu_S$ and $\sigma_S$ are the mean and standard
deviation of the \emph{flux density distribution}.
The probability of observing a pulsar above the limit $S_{\rm min}$ is
then
\begin{eqnarray}
\label{eq_p_obs}
p_{\rm obs} & = & \int_{\log S_{\rm min}}^{\infty} p(\log S)~d(\log S) \nonumber\\
& = & \frac{1}{2}~{\rm erfc}\left(\frac{\log S_{\rm min} - \mu_S}{\sqrt{2} \sigma_S}\right).
\end{eqnarray}

\begin{figure*}
\includegraphics[scale=0.35]{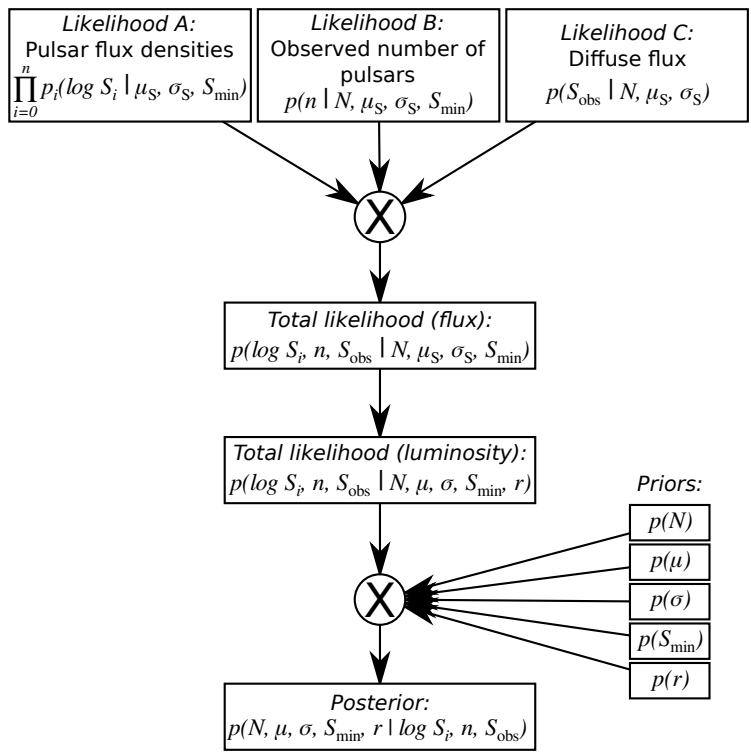}
\caption{Logical flow of the Bayesian analysis. The circle with the $\times$ sign
symbolizes a multiplication operation.
\label{fig_bayes_block}}
\end{figure*}

Our analysis involves computing three likelihoods in the flux domain based
on three sets of data, computing the total likelihood as the product of these
three likelihoods, converting this flux domain likelihood to the
luminosity domain, and subsequently, applying priors to obtain the posterior.
This procedure is depicted graphically in the block diagram of
Figure~\ref{fig_bayes_block}.

\subsection{Using pulsar flux densities}\label{subsec_step1}

In the first step, we consider as data the measured flux densities of the
pulsars in the cluster under scrutiny, that we call $\{S_i\}$. Ideally, the survey
sensitivity limit $S_{\rm min}$ can be taken as another datum, but its exact
value is not always known. The effects of radiometer noise, Doppler smearing,
interference, and in some cases, interstellar scintillation,
result in a distribution of $S_{\rm min}$. We decided, therefore,
to parametrize $S_{\rm min}$. The likelihood of observing a set of pulsars with
fluxes $\{S_i\}$ is represented as
\begin{equation*}
\prod\limits_{i=1}^{n}p_i(\log S_i | \mu_S, \sigma_S, S_{\rm min}),
\end{equation*}
where $n$ is the number of observed pulsars in the cluster. Each term in this product
is given by
\begin{equation}
p_i(\log S_i | \mu_S, \sigma_S, S_{\rm min}) =
\frac{1}{p_{\rm obs} \sigma_S \sqrt{2 \pi}}
e^{-\frac{(\log S_i - \mu_S)^2}{2 \sigma_S^2}},
\end{equation}
where $p_{\rm obs}$ is as given in Equation~(\ref{eq_p_obs}). This likelihood
is represented as `Likelihood A' in Figure~\ref{fig_bayes_block}. Uncertainties
in the flux density measurements are not considered here, but note that
ignoring them will have the effect of underestimating the width of posterior
credible intervals. In \S\ref{sec_qcu}, we discuss the effects of ignoring the
errors associated with flux density measurements.

\subsection{Incorporating the number of observed pulsars}

To infer the total number of pulsars $N$ in the cluster, we follow
\citet{boy11} to take as likelihood the probability of observing $n$ pulsars in
a cluster with $N$ pulsars, given by the binomial distribution
\begin{equation}
p(n | N, \mu_S, \sigma_S, S_{\rm min}) = \frac{N!}{n!(N-n)!}
~p_{\rm obs}^{n}~(1 - p_{\rm obs})^{N - n},
\end{equation}
where $p_{\rm obs}$ is computed as in Equation~(\ref{eq_p_obs}). This
likelihood is shown as `Likelihood B' in Figure~\ref{fig_bayes_block}.

\subsection{Considering diffuse emission}

Next, we incorporate information about the observed diffuse flux $S_{\rm obs}$
from the direction of the globular cluster. We assume that all radio emission
is due to the pulsars in the cluster, including both detected pulsars and
the unresolved background. Unlike standard pulsar searches,
imaging the diffuse radio emission of a cluster to estimate the number of
pulsars therein, is not affected by phenomena that cause pulse broadening,
such as dispersion or scattering, or the fact that some of the pulsars in the
cluster are in accelerating frames \citep{fru90}. For the likelihood of
measuring the diffuse flux $S_{\rm obs}$, we choose
\begin{equation}
\label{eq_likelihood3}
p(S_{\rm obs} | N, \mu_S, \sigma_S) = \frac{1}{\sigma_{\rm diff} \sqrt{2 \pi}}
~e^{-\frac{(S_{\rm obs} - S_{\rm diff})^2}{2 \sigma_{\rm diff}^2}},
\end{equation}
where $S_{\rm diff}$ is the expectation of the total diffuse flux of a cluster
whose flux density distribution is a log-normal with parameters $\mu_S$ and $\sigma_S$,
and having $N$ pulsars, and $\sigma_{\rm diff}$ is the standard deviation of this
distribution. This likelihood is referred to as `Likelihood C' in
Figure~\ref{fig_bayes_block}. Assuming that the cluster contains $N$ pulsars,
each having average luminosity,
\begin{equation}
S_{\rm diff} = N \langle S \rangle,
\end{equation}
and
\begin{equation}
\sigma_{\rm diff} = \sqrt{N}~{\rm SD}(S),
\end{equation}
where the expectation of $S$,
\begin{equation}
\langle S \rangle = 10^{\mu_S + \frac{1}{2} \sigma_S^2 \ln(10)},
\end{equation}
and the standard deviation of $S$,
\begin{equation}
{\rm SD}(S) = 10^{\mu_S + \frac{1}{2} \sigma_S^2 \ln(10)}
\sqrt{10^{\sigma_S^2 \ln(10)} - 1}.
\end{equation}
We do not consider the uncertainty in the diffuse flux measurement, and as
mentioned in \S\ref{subsec_step1}, this has the effect of underestimating the
credible intervals on our posteriors.

The total likelihood, then,
\begin{equation}
\begin{split}
p(\log S_i, &n, S_{\rm obs} | N, \mu_S, \sigma_S, S_{\rm min})\\
= &\prod\limits_{i=1}^{n}p_i(\log S_i | \mu_S, \sigma_S, S_{\rm min})\\
&\times p(n | N, \mu_S, \sigma_S, S_{\rm min})\\
&\times p(S_{\rm obs} | N, \mu_S, \sigma_S).
\end{split}
\label{eq_total_s_likelihood}
\end{equation}

\subsection{Transformation to luminosity domain}

The flux density distribution of pulsars in a globular cluster, although
proportional to their luminosity distribution, is not suitable for comparing
the populations in different clusters, as it depends on the distance to the
cluster. It is, therefore, useful to transform the total likelihood obtained
in the previous subsection to the luminosity domain. We convert the total
likelihood of Equation~(\ref{eq_total_s_likelihood}) to the luminosity domain
in the following way. Since the pseudo-luminosity equation can be written in terms of
logarithms as $\log L = \log S + 2 \log r$, where $L$ is in mJy kpc$^2$, $S$ is in
mJy, and $r$ is in kpc, the means of the two distributions are related additively
by the term $2 \log r$, while the standard deviations are the same. Taking into
account the uncertainty in distance, we have a distribution of distances,
$p(r)$. The total likelihood in the luminosity domain is
\begin{equation}
\begin{split}
p(\log S_i, &n, S_{\rm obs} | N, \mu, \sigma, S_{\rm min}, r)\\
&= p(\log S_i, n, S_{\rm obs} | N, \mu_S, \sigma_S, S_{\rm min}),
\end{split}
\end{equation}
where $\mu$ and $\mu_S$ are related additively as mentioned above, and $\sigma$
and $\sigma_S$ are equal. The final joint posterior in luminosity is then
given by
\begin{equation}
\begin{split}
p(N, &\mu, \sigma, S_{\rm min}, r | \log S_i, n, S_{\rm obs})\\
\propto~&p(\log S_i, n, S_{\rm obs} | N, \mu, \sigma, S_{\rm min}, r)\\
&\times p(N)~p(\mu)~p(\sigma)~p(S_{\rm min})~p(r).
\end{split}
\end{equation}
The prior on $N$ is taken to be uniform from $n$ to $\infty$, where
the upper limit, for the sake of computation, would be a sufficiently large value.
We also use uniform priors on the model parameters $\mu$ and $\sigma$.
Due to the nature of the uncertainty in determining the exact value of
$S_{\rm min}$, we choose a uniform prior on it in the range (0, min($S_i$)],
where the upper limit is the flux density of the least bright pulsar in the
cluster. The prior on $r$ is taken to be a Gaussian. This joint posterior is
then integrated over various sets of model parameters to obtain 
marginalized posteriors.

\section{Applications}\label{sec_apps}

\begin{figure*}
\includegraphics[angle=-90,width=\textwidth]{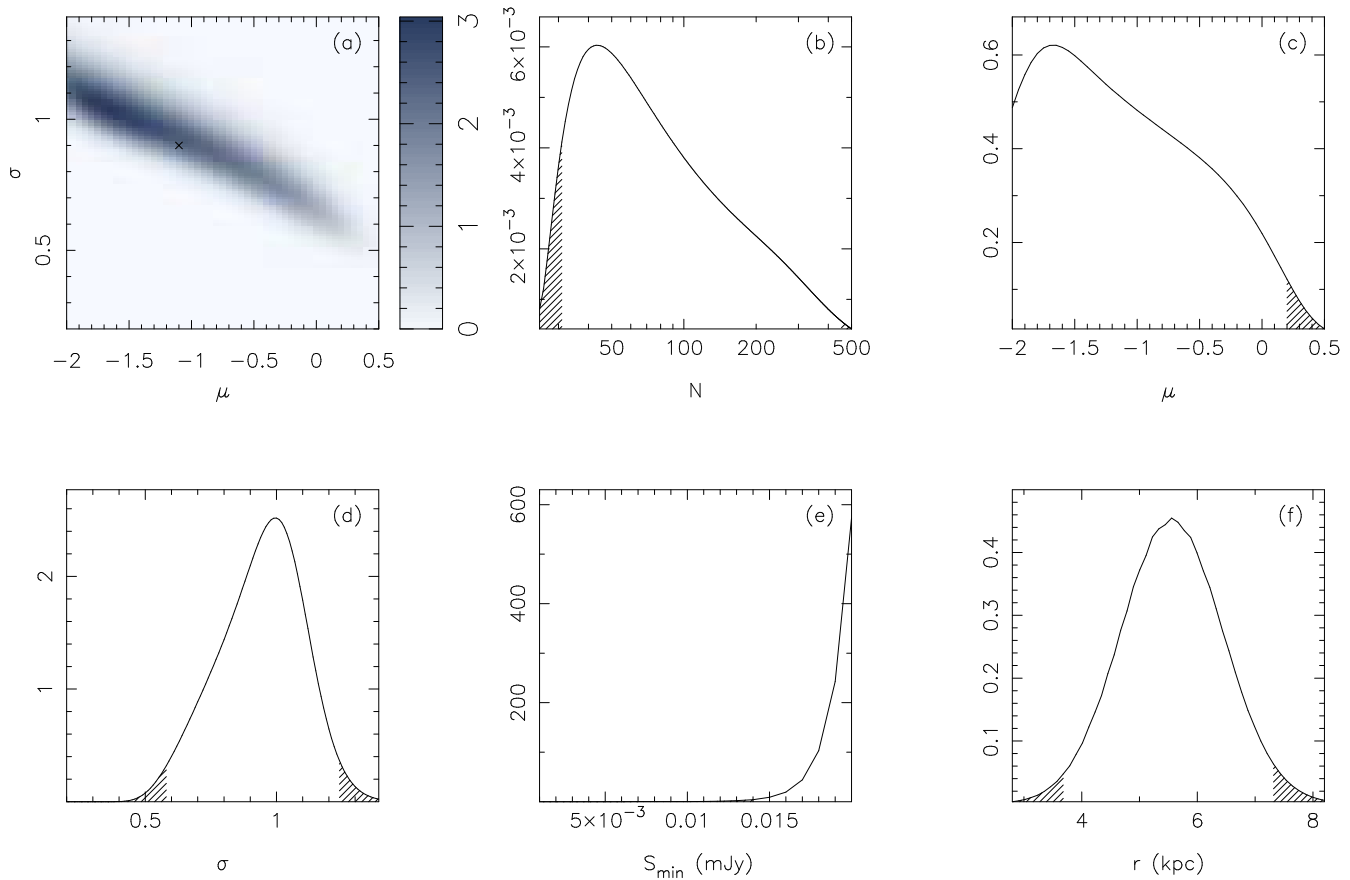}
\caption{Results of the Bayesian analysis for Terzan 5, with $\{S_i\}$ as given in
Table~\ref{tab_flux}, $n = 25$, and $S_{\rm obs} = 5.2$ mJy. This analysis was
run with wide priors on $\mu$ and $\sigma$, with the ranges equal to those used
by \citet{bag11} (their Figure 2). (a) depicts the joint posterior on $\mu$ and $\sigma$,
marginalized over $N$, $S_{\rm min}$ and $r$. The `$\times$' symbol marks the intersection
of the values obtained by \citet{fau06}; (b) is the
marginalized posterior for $N$, with
a mode of 43 and a median of 142. The x-axis is plotted in log scale for
clarity; (c) is the marginalized posterior for $\mu$ with a mode of $-$1.65
and a median of $-$1.2; (d) is the
marginalized posterior for $\sigma$
with a mode of 1.0 and a median of 0.95; (e) is the marginalized posterior for
$S_{\rm min}$  with both mode and
median equal to 0.02 mJy; (f) is the marginalized posterior for $r$, with both
mode and median equal to 5.56 kpc. The hatching indicates regions that lie
outside a 95\% credible interval.
\label{fig_ter5_wide}}
\end{figure*}

We applied our Bayesian technique\footnote{The software package that we
developed to perform the analysis described in this paper is available freely
for download from http://psrpop.phys.wvu.edu/gcbayes/.} to the globular
clusters Terzan 5, 47 Tucanae and M 28 (although the
clusters we consider contain only recycled known pulsars, the analysis would remain
the same even if there were young pulsars in the sample). The choice of
clusters was based on the amount of data available. Terzan 5 is the cluster most-suited for
this analysis due to the fact that it has a relatively large number of pulsars
for which flux density measurements are available. 
Although Terzan 5 has 34 known pulsars (Ransom S. M., private communication), we take $n = 25$, the
number of pulsars for which we have flux density measurements. The flux
densities of the individual pulsars were collected in a literature survey
\citep[][and references therein]{bag11}, with the values relevant to this work
reproduced in Table~\ref{tab_flux}. For Terzan 5, the flux densities we used
were scaled from those reported at 1950 MHz by \citet{ran05} and \citet{hes06}
to 1400 MHz using a spectral index, $\alpha = -1.9$ (the mean
value for globular cluster MSPs), using the power law
$S(\nu) \propto \nu^\alpha$. \citet{hes07} and \citet{bag11} discuss the choice
of spectral index in detail. The observed diffuse flux density at 1400 MHz is
taken to be $S_{\rm obs} = 5.2$ mJy \citep[the sum of the diffuse flux and the
fluxes of point sources as reported by][]{fru00}. The priors used were formed
in the following ways. The prior on $N$ was chosen to be a uniform distribution
in the range [$n$, 500], where the upper limit is 150\% of the upper limit obtained by
\citet{bag11} \citep[using the values of $\mu$ and $\sigma$ as found by][]{fau06}
above their upper limit. We note that this prior is sufficiently wide to ensure
that the posterior does not rail against the prior boundaries.
We chose uniform distributions in the same range of $\mu$ and
$\sigma$ as used by \citet{bag11} as our priors, namely, [-2.0, 0.5] and
[0.2, 1.4], respectively. Survey sensitivity limits were
not always available, and additionally, due to a variety of factors mentioned
in \S\ref{subsec_step1}, for all of our analyses, we took $S_{\rm min}$ to be
a uniform distribution in the range (0, min($S_i$)]. The most recent
measurement of the distance to Terzan 5, $r = 5.5\pm0.9$ kpc
\citep{ort07}, was used to model the distance prior. We modelled the distance
prior as a Gaussian with mean 5.5 kpc and standard deviation 0.9 kpc.
Figure~\ref{fig_ter5_wide} shows the results of the analysis for Terzan 5.
The mode of the marginalized posterior on $N$, shown in Figure~\ref{fig_ter5_wide}(b),
is 43 and the median with the surrounding 95\% credible interval is $142^{+310}_{-110}$.
As can be seen from Figures~\ref{fig_ter5_wide}(b), \ref{fig_ter5_wide}(c) and
\ref{fig_ter5_wide}(d), the constraints we obtain on $N$, $\mu$ and $\sigma$,
respectively, are broad, due to a dearth of flux density
measurements. The marginalized posterior on $S_{\rm min}$, plotted in
Figure~\ref{fig_ter5_wide}(e), shows a strong preference for values away from 0
and closer to that of the least bright pulsar observed. The main results are
tabulated in Table~\ref{tab_bayes}.

For 47 Tucanae and M 28, containing 14 and 9 pulsars each, the individual flux
densities used are given in Table~\ref{tab_flux}. We took $S_{\rm obs} = 2.0$ mJy
\citep[1400 MHz flux as reported by][]{mcc04} for 47 Tucanae, and $S_{\rm obs} = 1.8$ mJy
\citep[1400 MHz flux as reported by][]{kul90a} for M 28. The priors on $N$ were
taken to be uniform in the intervals [$n$, 225] for 47 Tucanae and [$n$, 400]
for M 28, where the upper limits were computed in the same way as we did for Terzan 5. Priors on
$S_{\rm min}$ were formed as in the case of Terzan 5, in the range
(0, min($S_i$)]. The latest distance measurement of $4.69\pm0.17$ kpc \citep{woo12}
was used to form the distance prior for 47 Tucanae. For M 28, we used
$r = 5.5\pm0.3$ kpc \citep{ser12}. The main results for these
clusters are tabulated in Table~\ref{tab_bayes}.

\begin{table}
\begin{minipage}{80mm}
\addtocounter{footnote}{-1}
\footnotetext[\value{footnote}]{Based on values reported by \citet{ran05}. The
fractional uncertainties on these values are $\sim 30\%$.}
\addtocounter{footnote}{1}
\footnotetext[\value{footnote}]{Based on the value reported by \citet{hes06}. The
fractional uncertainty on this value is $25\%$.}
\addtocounter{footnote}{1}
\footnotetext[\value{footnote}]{\citet{cam00}. The fractional uncertainties on these
values range from $10\%$ to $40\%$.}
\addtocounter{footnote}{1}
\footnotetext[\value{footnote}]{\citet{beg06}. The reported values are `highly
uncertain'.}
\addtocounter{footnote}{-3}
\begin{center}
\caption{Flux densities used in the analysis.\label{tab_flux}}
\begin{tabular}{lrrrr}
\hline
Pulsar & 1400 MHz Flux Density (mJy)\\
\hline
Terzan 5\\
\hline
J1748$-$2446A & 1.91$^{\aaalph{footnote}}$\\
J1748$-$2446C & 0.68$^{\aaalph{footnote}}$\\
J1748$-$2446D & 0.08$^{\aaalph{footnote}}$\\
J1748$-$2446E & 0.09$^{\aaalph{footnote}}$\\
J1748$-$2446F & 0.07$^{\aaalph{footnote}}$\\
J1748$-$2446G & 0.03$^{\aaalph{footnote}}$\\
J1748$-$2446H & 0.03$^{\aaalph{footnote}}$\\
J1748$-$2446I & 0.05$^{\aaalph{footnote}}$\\
J1748$-$2446J & 0.04$^{\aaalph{footnote}}$\\
J1748$-$2446K & 0.08$^{\aaalph{footnote}}$\\
J1748$-$2446L & 0.08$^{\aaalph{footnote}}$\\
J1748$-$2446M & 0.06$^{\aaalph{footnote}}$\\
J1748$-$2446N & 0.10$^{\aaalph{footnote}}$\\
J1748$-$2446O & 0.23$^{\aaalph{footnote}}$\\
J1748$-$2446P & 0.14$^{\aaalph{footnote}}$\\
J1748$-$2446Q & 0.05$^{\aaalph{footnote}}$\\
J1748$-$2446R & 0.02$^{\aaalph{footnote}}$\\
J1748$-$2446S & 0.03$^{\aaalph{footnote}}$\\
J1748$-$2446T & 0.04$^{\aaalph{footnote}}$\\
J1748$-$2446U & 0.03$^{\aaalph{footnote}}$\\
J1748$-$2446V & 0.13$^{\aaalph{footnote}}$\\
J1748$-$2446W & 0.04$^{\aaalph{footnote}}$\\
J1748$-$2446X & 0.03$^{\aaalph{footnote}}$\\
J1748$-$2446Y & 0.03$^{\aaalph{footnote}}$\addtocounter{footnote}{1}\\
J1748$-$2446ad & 0.15$^{\aaalph{footnote}}$\addtocounter{footnote}{1}\\
\hline
47 Tucanae\\
\hline
J0023$-$7204C & 0.36$^{\aaalph{footnote}}$\\
J0024$-$7204D & 0.22$^{\aaalph{footnote}}$\\
J0024$-$7205E & 0.21$^{\aaalph{footnote}}$\\
J0024$-$7204F & 0.15$^{\aaalph{footnote}}$\\
J0024$-$7204G & 0.05$^{\aaalph{footnote}}$\\
J0024$-$7204H & 0.09$^{\aaalph{footnote}}$\\
J0024$-$7204I & 0.09$^{\aaalph{footnote}}$\\
J0023$-$7203J & 0.54$^{\aaalph{footnote}}$\\
J0024$-$7204L & 0.04$^{\aaalph{footnote}}$\\
J0023$-$7205M & 0.07$^{\aaalph{footnote}}$\\
J0024$-$7204N & 0.03$^{\aaalph{footnote}}$\\
J0024$-$7204O & 0.10$^{\aaalph{footnote}}$\\
J0024$-$7204Q & 0.05$^{\aaalph{footnote}}$\\
J0024$-$7203U & 0.06$^{\aaalph{footnote}}$\addtocounter{footnote}{1}\\
\hline
M 28\\
\hline
B1821$-$24A & 0.94$^{\aaalph{footnote}}$\\
J1824$-$2452B & 0.07$^{\aaalph{footnote}}$\\
J1824$-$2452C & 0.17$^{\aaalph{footnote}}$\\
J1824$-$2452D & 0.05$^{\aaalph{footnote}}$\\
J1824$-$2452E & 0.06$^{\aaalph{footnote}}$\\
J1824$-$2452F & 0.08$^{\aaalph{footnote}}$\\
J1824$-$2452G & 0.05$^{\aaalph{footnote}}$\\
J1824$-$2452H & 0.06$^{\aaalph{footnote}}$\\
J1824$-$2452J & 0.07$^{\aaalph{footnote}}$\\
\hline
\end{tabular}
\end{center}
\end{minipage}
\end{table}

The value of $N$ can be further refined by considering possible dependences
on other physical parameters of globular clusters. In a forthcoming paper
\citep{tur13}, an empirical Bayesian approach is being applied to the set of 95
flux density limits for globular clusters presented in \citet{boy11} in which
pulsar abundance as a function of two-body encounter rate, metallicity, cluster
mass, etc. is incorporated into the likelihood functions. Note that $N$ is the
size of the population of pulsars in the cluster that are
beaming towards the Earth. We can include the beaming fraction -- the fraction of
all pulsars beaming towards us -- to refine this estimate. Uncertainties notwithstanding,
the beaming fraction of millisecond pulsars is generally thought to be greater than 50\%
\citep{kra98}. This, together with the fact that most pulsars in globular clusters
are millisecond pulsars, imply that the true population size in a cluster is approximately
a factor of two more than the potentially observable population size.

\subsection{Using prior information}

\begin{figure*}
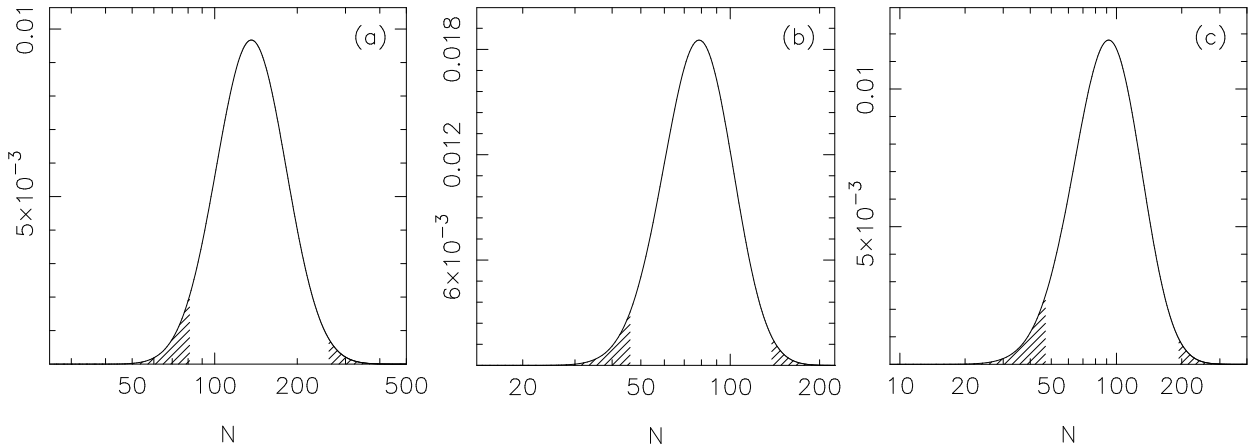

\subfloat{
\includegraphics[angle=-90,width=0.315\textwidth]{ter5_narrow.ps}}
\subfloat{
\includegraphics[angle=-90,width=0.3045\textwidth]{47tuc_narrow.ps}}
\subfloat{
\includegraphics[angle=-90,width=0.3\textwidth]{m28_narrow.ps}}
\caption{Posteriors on $N$ after applying the \citet{boy11} priors on
$\mu$ and $\sigma$: (a) Terzan 5, with $n$ = 25 and $S_{\rm obs} = 5.2$ mJy.
The median with the surrounding 95\% credible interval of $N$ is
$147^{+112}_{-65}$; (b) 47 Tucanae, with $n = 14$ and $S_{\rm obs} = 2.0$ mJy.
The median with credible interval is $83^{+54}_{-35}$; (c) M 28, with $n = 9$
and $S_{\rm obs} = 1.8$ mJy. The median with credible interval is
$100^{+91}_{-52}$. The flux densities of the individual pulsars, $\{S_i\}$, used
in this analysis are given in Table~\ref{tab_flux}.
\label{fig_prior}}
\end{figure*}

In the framework developed in the previous section, we use broad uniform
(non-informative) priors for the mean and standard deviation of the
log-normal. This lack of prior information is apparent in
Figure~\ref{fig_ter5_wide}(b), where $N$ is not very well constrained.
Prior information can help better constrain the parameters of interest.
\citet{boy11} use models of non-recycled Galactic pulsars from \citet{rid10}
to narrow down $\mu$ to between $-$1.19 and $-$1.04, and $\sigma$ to the range
0.91 to 0.98. We assume that these values are applicable to the
globular cluster pulsars based on \citet{bag11} who draw the conclusion that
the luminosity function of cluster pulsars is no different from that of the
Galactic disc as found by \citet{fau06}. The values themselves are also consistent with the results of
\citet{bag11}. We choose $\mu$ and $\sigma$ to be uniform within these ranges.
Applying the Bayesian analysis over this narrower range of $\mu$ and $\sigma$
for Terzan 5 results in much tighter constraints on $N$ as seen in Figure~\ref{fig_prior}(a),
in which the mode of the distribution is 136 and the median and a 95\% credible interval
is $147^{+112}_{-65}$. The analysis was also performed for 47 Tucanae and M 28,
the results of which are given in Figures~\ref{fig_prior}(b) and \ref{fig_prior}(c), respectively.
For 47  Tucanae, the mode of $N$ is 79 and the median with the surrounding 95\%
credible interval is $83^{+54}_{-35}$. For M 28, the mode is 91 and the median
with credible interval is $100^{+91}_{-52}$. Our result for Terzan 5 is
consistent with that of \citet{bag11}. In the case of 47 Tucanae and M 28, there
is partial, yet considerable overlap between our credible ranges and the
corresponding confidence intervals of \citet{bag11}. For 47 Tucanae, our
results agree with those of \citet{gri02}, i.e., 35--90 MSPs with X-ray
luminosities above $10^{30}$~erg~s$^{-1}$. However, our result
for 47 Tucanae is inconsistent with that of \citet{mcc04} who estimate
$N \leq 30$. This disparity may be due to the high scintillation of the
pulsars in this cluster affecting both individual as well as diffuse
flux measurements. The results of our analyses are tabulated in
Table~\ref{tab_bayes}.

\begin{table}
\caption{Median and 95\% credible intervals from the various analyses presented
in this paper. We note here that, in addition to the sources of error mentioned
in the text, the values of $\mu$ and $\sigma$ presented here are also affected
by the fact that computations are discrete and hence use a finite number of
steps. Note that for the case of narrow priors on $\mu$ and $\sigma$, the
corresponding two columns do not carry any useful information, reflecting merely
the ranges of the priors, and are included here only for completeness.
\label{tab_bayes}}
\begin{center}
\begin{tabular}{lrrr}
\hline
Cluster & $N$ & $\mu$ & $\sigma$\\
\hline
Wide priors on $\mu$ and $\sigma$\\
\hline
Ter 5 & $142^{+310}_{-110}$ & $-1.2^{+1.4}_{-0.8}$ & $1.0^{+0.3}_{-0.4}$\\
47 Tuc & $39^{+169}_{-25}$ & $-0.6^{+0.9}_{-1.3}$ & $0.7^{+0.4}_{-0.4}$\\
M 28 & $198^{+191}_{-169}$ & $-1.3^{+1.1}_{-0.7}$ & $0.8^{+0.3}_{-0.3}$\\
\hline
Narrow priors on $\mu$ and $\sigma$\\
\hline
Ter 5 & $147^{+112}_{-65}$ & $-1.12^{+0.08}_{-0.07}$ & $0.94^{+0.03}_{-0.03}$\\
47 Tuc & $83^{+54\hphantom{0}}_{-35}$ & $-1.13^{+0.08}_{-0.07}$ & $0.94^{+0.04}_{-0.03}$\\
M 28 & $100^{+91\hphantom{0}}_{-52}$ & $-1.13^{+0.09}_{-0.06}$ & $0.94^{+0.04}_{-0.03}$\\
\hline
\end{tabular}
\end{center}
\end{table}

\section{Discussion}

\subsection{Effect of flux density measurement errors on credible intervals} \label{sec_qcu}

Our analysis in its present form does not take into account measurement
uncertainties of pulsar flux densities. This leads to an underestimation of our
credible intervals and in this section, we discuss the effect that neglecting
measurement errors has on our credible intervals. We performed a Monte Carlo
simulation in which the flux density corresponding to each Terzan 5 pulsar was
modelled as a Gaussian with mean equal to the measured value (given in Table~\ref{tab_flux})
and standard deviation equal to the measurement error (given in the footnotes to
Table~\ref{tab_flux}). A flux density value was picked for
each pulsar from these distributions and our Bayesian analysis performed on the
new set of flux densities. The analysis was done with both wide and narrow
priors on $\mu$ and $\sigma$, resulting in two sets of credible intervals. For
non-informative priors, the standard deviation on the lower limit of the
credible interval for $N$ was found to be 7 while that on the upper limit was
165. For narrow priors, the standard deviation on the lower limit was 19 while
that on the upper limit was 75. This simplified simulation of the impact of
unmodelled measurement uncertainties suggests that the lower limits of the 95\%
credible intervals are fairly robust, while their upper limits may vary by
about one half of the values given in Table~\ref{tab_bayes}. A more accurate
simulation would involve generating sets of flux densities according to all
the priors in our analysis, with added dither due to the unmodelled measurement
uncertainty, and for each set, compute the 95\% credible intervals using our
technique, and check what fraction of true values lies outside these intervals.
Such a simulation, although more accurate, would be computationally expensive.
In principle, measurement uncertainties could be included directly in the
likelihood model, and marginalized over to compute the posteriors of interest.

\subsection{Effect of increasing detections on credible intervals} \label{sec_fracpop}

\begin{figure*}
\begin{tabular}{c}
\subfloat{
\includegraphics[angle=0,width=0.33\textwidth]{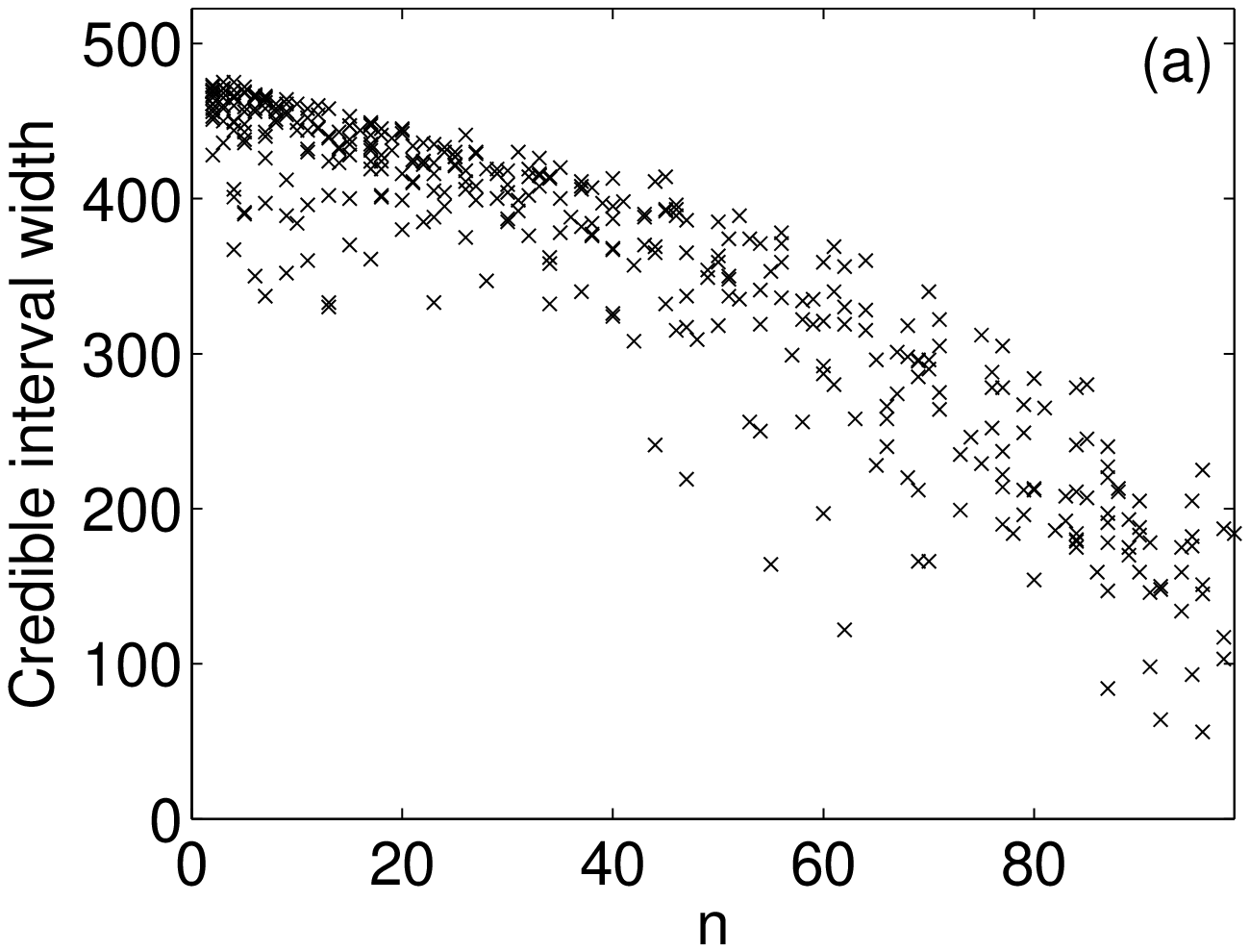}}
\subfloat{
\includegraphics[angle=0,width=0.33\textwidth]{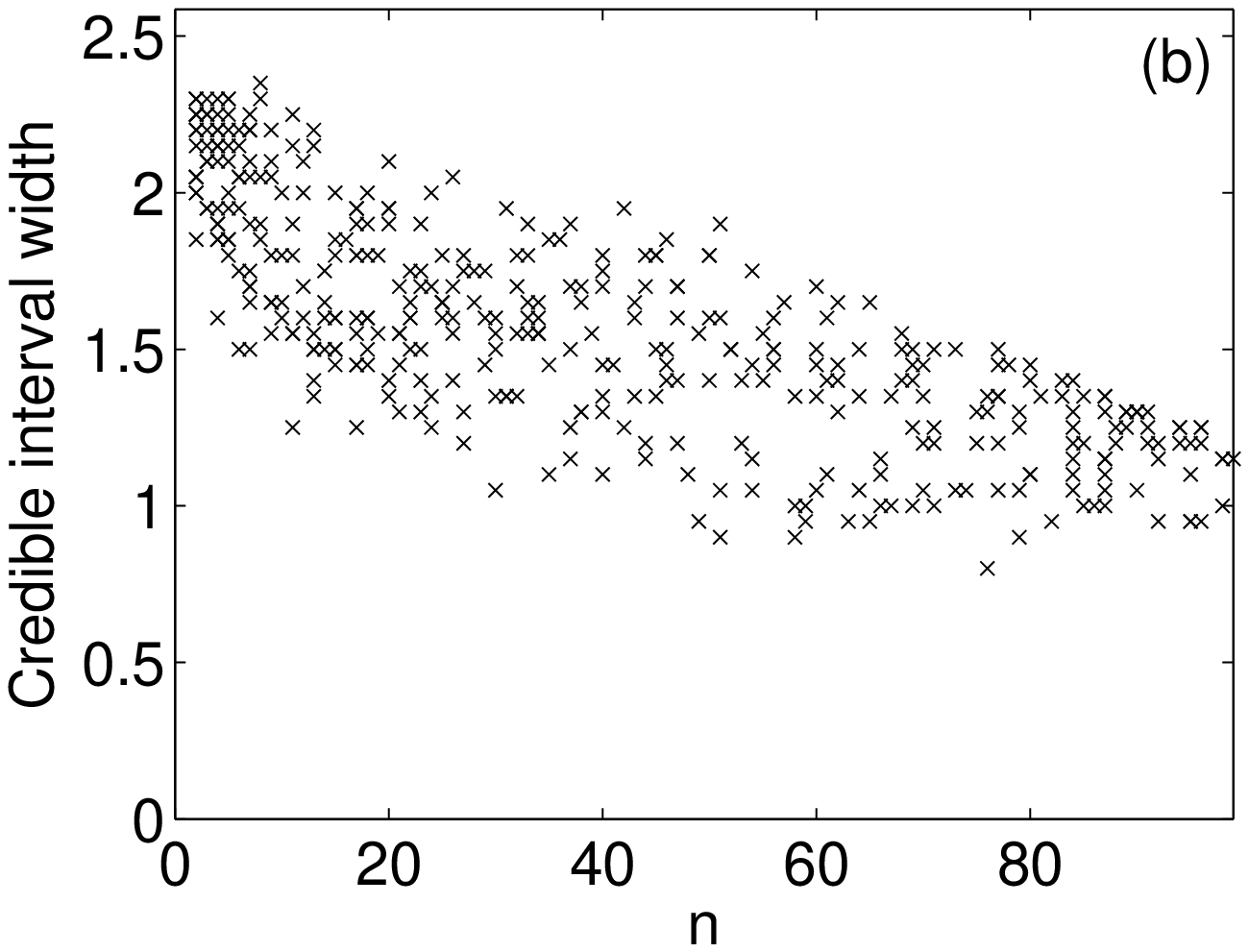}}
\subfloat{
\includegraphics[angle=0,width=0.33\textwidth]{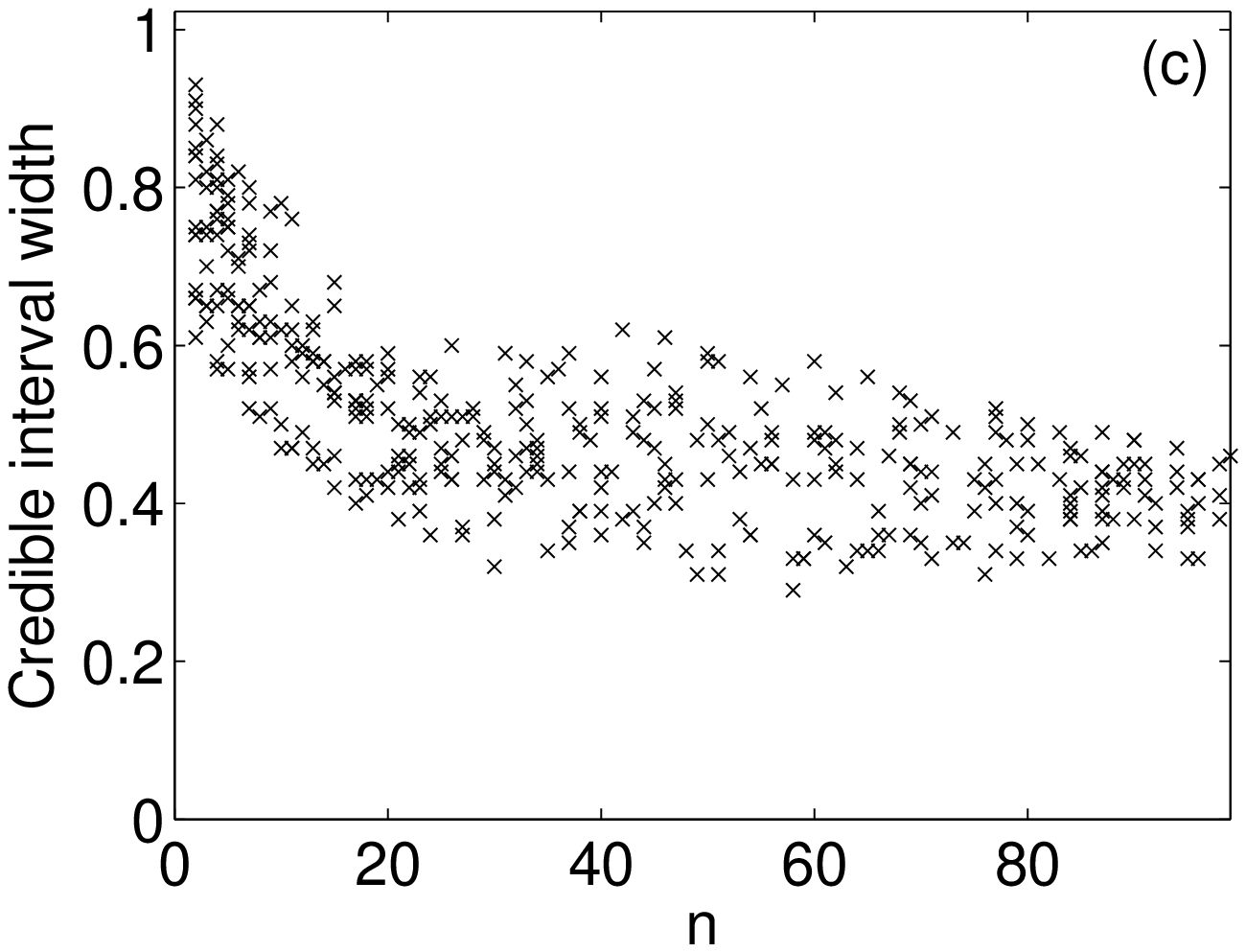}}
\end{tabular}
\caption{Results of the Monte Carlo simulation described in
\S\ref{sec_fracpop}. (a) shows how the width of the credible interval on $N$
decreases with increasing number of detected pulsars, $n$. (b) and (c)
correspond to credible intervals on $\mu$ and $\sigma$, respectively.
\label{fig_fracpop}}
\end{figure*}

In order to gauge the performance of our technique with respect to
increasing pulsar detections and subsequent flux density measurements, we
performed the following Monte Carlo simulation. We simulated a globular
cluster with population size equal to our wide-prior median estimate for
Terzan 5, 142 pulsars, located at the distance of Terzan 5, whose luminosity
follows a log-normal with $\mu$ and $\sigma$ fixed at the \citet{fau06} values
of $-1.1$ and $0.9$, respectively. For this cluster, we varied the survey
sensitivity limit, and at each step, counted the observed number of pulsars,
and ran our Bayesian analysis, giving us a set of credible intervals. The
Bayesian analysis was done with the same priors as in the first part of
\S\ref{sec_apps}, viz. uniform prior on $N$ in the range [$n$, 500],
uniform prior on $S_{\rm min}$ in the range (0, min($S_i$)], and the uniform,
wide \citet{bag11} priors on $\mu$ and $\sigma$. This process was then repeated
for multiple Monte Carlo realizations of the log-normal to allow for flux
density variations to be manifested. The results are given in
Figure~\ref{fig_fracpop} where the width of the credible intervals on the
parameters $N$, $\mu$ and $\sigma$ are plotted against the number of pulsars
detected. As expected, there is a clear improvement in the credible interval
widths with the number of pulsars. Since our population estimate of 142
indicates that we have already detected about a fifth of the potentially
observable pulsars in Terzan 5, increasing the number of detections/flux
density measurements by, say, a factor of 2 would improve our credible interval
on $N$ by approximately 15\%, whereas the credible intervals on $\mu$ an
$\sigma$ would improve by about 15\% and 10\%, respectively.

\section{Summary and conclusions} \label{sec_conclusions}

We have developed a Bayesian technique to constrain the luminosity function parameters
and population size of pulsars in individual globular clusters, given a data
set that consists of the number of observed pulsars, the flux densities of the
individual pulsars in the cluster and the total diffuse flux emission from the
direction of the globular cluster, assuming a log-normal luminosity function. We have
applied our analysis to a few globular clusters and have demonstrated the
utility of this technique in constraining the aforementioned parameters.

Our technique is applied in two different ways -- first, with no prior
information, and second, assuming prior knowledge of the possible ranges of
$\mu$ and $\sigma$. As shown for Terzan 5, the results for the first
approach do not constrain $N$, $\mu$ or $\sigma$ very well due to paucity of
data, but the latter two do exhibit consistency with the values found by
\citet{fau06} and \citet{bag11}. For the second approach in which we assume
prior information to bound $\mu$ and $\sigma$, the priors help better constrain
the total number of pulsars in the cluster.

The technique we have developed here should prove useful in future studies of
the globular cluster luminosity function where ongoing and future pulsar surveys
are expected to provide a substantial increase in the observed populations of pulsars
in many clusters. In particular, we anticipate that the increased amount of data
would enable us to constrain the distributions of $\mu$ and $\sigma$ independently
(i.e. without the need to assume prior information from the Galactic pulsar population).
Further interferometric measurements of the diffuse radio flux in many globular
clusters could provide improved constraints on $\mu$ and $\sigma$ by measuring
the flux contribution from the individually unresolvable population of pulsars.

\section*{Acknowledgments}

We thank the anonymous referee for useful comments. We thank Phil Turk and
Nipuni Palliyaguru for useful discussions. This work was supported
by a Research Challenge Grant to the WVU Center for Astrophysics by the West
Virginia EPSCoR foundation, and also by the Astronomy and Astrophysics Division
of the National Science Foundation via a grant AST-0907967.


\begin{thebibliography}{}
\bibitem[\protect\citeauthoryear{Alpar et al.}{1982}]{alp82}Alpar M. A.,
Cheng A. F., Ruderman M. A., Shaham J., 1982, Nat, 300, 728
\bibitem[\protect\citeauthoryear{Archibald et al.}{2009}]{arc09}Archibald A. M.,
Stairs I. H., Ransom S. M., Kaspi V. M., Kondratiev V. I., Lorimer D. R.,
McLaughlin M. A, Boyles J., Hessels J. W. T., Lynch R., van Leeuwen J.,
Roberts M. S. E., Jenet F., Champion D. J., Rosen R., Barlow B. N.,
Dunlap B. H., Remillard R. A., 2009, Sci, 324, 1411
\bibitem[\protect\citeauthoryear{Bagchi, Lorimer \& Chennamangalam}
{Bagchi et al.}{2011}]{bag11}Bagchi M., Lorimer D. R., Chennamangalam J., 2011,
MNRAS, 418, 477
\bibitem[\protect\citeauthoryear{B\'egin}{2006}]{beg06}B\'egin S., 2006,
Ph.D. Thesis, UBC
\bibitem[\protect\citeauthoryear{Boyles et al.}{2011}]{boy11}
Boyles J., Lorimer D. R., Turk P. J., Mnatsakanov R., Lynch R. S.,
Ransom S. M., Freire P. C., Belczynski K., 2011, ApJ, 742, 51
\bibitem[\protect\citeauthoryear{Camilo \& Rasio}{2005}]{cam05}Camilo F.,
Rasio F. A., 2005, in Rasio F. A., Stairs I. H., eds, ASP Conf. Ser. Vol. 328,
Binary Radio Pulsars. Astron. Soc. Pac., San Francisco, p.147
\bibitem[\protect\citeauthoryear{Camilo et al.}{2000}]{cam00}Camilo, F.,
Lorimer D. R., Freire P., Lyne A. G., Manchester R. N., 2000, ApJ,
535, 975
\bibitem[\protect\citeauthoryear{Faucher-Gigu\`ere \& Kaspi}{2006}]{fau06}
Faucher-Gigu\`ere C.-A., Kaspi V. M., 2006, ApJ, 643, 332
\bibitem[\protect\citeauthoryear{Fruchter \& Goss}{1990}]{fru90}
Fruchter A. S., Goss W. M., 1990, ApJ, 365, L63
\bibitem[\protect\citeauthoryear{Fruchter \& Goss}{2000}]{fru00}
Fruchter A. S., Goss W. M., 2000, ApJ, 536, 865
\bibitem[\protect\citeauthoryear{Gregory}{2005}]{gre05}Gregory P. C., 2005,
Bayesian Logical Data Analysis for the Physical Sciences: A Comparative
Approach with \emph{Mathematica} Support. Cambridge University Press,
Cambridge, UK
\bibitem[\protect\citeauthoryear{Grindlay et al.}{2002}]{gri02}Grindlay J. E.,
Camilo F., Heinke C. O., Edmonds P. D., Cohn H., Lugger P., 2002, ApJ, 581, 470
\bibitem[\protect\citeauthoryear{Hessels et al.}{2007}]{hes07}
Hessels J. W. T., Ransom S. M., Stairs I. H., Kaspi V. M.,
Freire P. C. C., 2007, ApJ, 670, 363
\bibitem[\protect\citeauthoryear{Hessels et al.}{2006}]{hes06}
Hessels J. W. T., Ransom S. M., Stairs I. H., Freire P. C. C., Kaspi V. M.,
Camilo F., 2006, Sci, 311, 1901
\bibitem[\protect\citeauthoryear{Kulkarni et al.}{1990}]{kul90a}Kulkarni S. R.,
Goss W. M., Wolszczan A., Middleditch J., 1990, ApJ, 363, L5
\bibitem[\protect\citeauthoryear{Kulkarni, Narayan \& Romani}
{Kulkarni et al.}{1990}]{kul90b}Kulkarni S. R., Narayan R., Romani R. W., 1990,
ApJ, 356, 174
\bibitem[\protect\citeauthoryear{Kramer et al.}{1998}]{kra98}Kramer M.,
Xilouris K. M., Lorimer D. R., Doroshenko O., Jessner A., Wielebinski R., 
Wolszczan A., Camilo F., 1998, ApJ, 501, 270
\bibitem[\protect\citeauthoryear{Lorimer \& Kramer}{2005}]{lor05}Lorimer D. R.,
Kramer M., 2005, Handbook of Pulsar Astronomy, Cambridge Univ. Press,
Cambridge, UK
\bibitem[\protect\citeauthoryear{McConnell et al.}{2004}]{mcc04}McConnell D.,
Deshpande A. A., Connors T., Ables J. G., 2004, MNRAS, 348, 1409
\bibitem[\protect\citeauthoryear{Meylan \& Heggie}{1997}]{mey97}Meylan G.,
Heggie D. C., 1997, A\&AR, 8, 1
\bibitem[\protect\citeauthoryear{Ortolani et al.}{2007}]{ort07}
Ortolani S., Barbuy B., Bica E., Zoccali M., Renzini A., 2007, A\&A, 470, 1043
\bibitem[\protect\citeauthoryear{Ransom et al.}{2005}]{ran05}Ransom S. M.,
Hessels J. W. T., Stairs I. H., Freire P. C. C., Camilo F., Kaspi V. M.,
Kaplan D. L., 2005, Sci, 307, 892
\bibitem[\protect\citeauthoryear{Ridley \& Lorimer}{2010}]{rid10}Ridley J. P.,
Lorimer D. R., 2010, MNRAS, 404, 1081
\bibitem[\protect\citeauthoryear{Servillat et al.}{2012}]{ser12}
Servillat M., Heinke C. O., Ho W. C. G., Grindlay J. E., Hong J., van den Berg M.,
Bogdanov S., 2012, MNRAS, 423, 1556
\bibitem[\protect\citeauthoryear{Turk \& Lorimer}{2013, in prep.}]{tur13}Turk P. J.,
Lorimer D. R., 2013, in prep.
\bibitem[\protect\citeauthoryear{Wall \& Jenkins}{2003}]{wal03}Wall J. V.,
Jenkins C. R., 2003, Practical Statistics for Astronomers. Cambridge Univ.
Press, Cambridge, UK
\bibitem[\protect\citeauthoryear{Woodley et al.}{2012}]{woo12}Woodley K. A.,
Goldsbury R., Kalirai J. S., Richer H. B., Tremblay P.-E., Anderson J., Bergeron P.,
Dotter A., Esteves L., Fahlman G. G., Hansen B. M. S., Heyl J., Hurley J., Rich R. M.,
Shara M. M., Stetson P. B., 2012, AJ, 143, 50
\bibitem[\protect\citeauthoryear{Zepf}{2003}]{zep03}
Zepf S. E., 2003, in Engvold O., ed, Highlights of Astronomy, Vol. 13,
as presented at the XXVth General Assembly of the IAU 2003,
Astron. Soc. Pac., San Francisco, p. 347
\end{thebibliography}
\end{document}